\documentstyle[twocolumn,aps,epsfig,prl,floats]{revtex}

\def \invpb    {\relax\ifmmode{\rm pb^{-1}}\else{$\rm pb^{-1}$}\fi}
\def \stop     {\relax\ifmmode{\tilde{t}}\else{$\tilde{t}$}\fi}
\def \ET       {\relax\ifmmode{{{\rm E}_{\scriptscriptstyle\rm T}}}
                \else{${{\rm E}_{\scriptscriptstyle\rm T}}$}\,\fi}
\def \sET       {\relax\ifmmode{{{\scriptstyle\rm E}_{\scriptscriptstyle\rm T}}}
                \else{${{\scriptstyle\rm E}_{\scriptscriptstyle\rm T}}$}\,\fi}
\def \MET      {\relax\ifmmode{\mbox{$\raisebox{.3ex}{$\not$}\ET$}}
                \else{$\mbox{$\raisebox{.3ex}{$\not$}\sET$}$}\fi}
\def \sMET      {\relax\ifmmode{\mbox{$\raisebox{.3ex}{$\scriptstyle\not$}\sET$}}
                \else{$\mbox{$\raisebox{.3ex}{$\not$}\ET$}$}\fi}
\def \mgev     {GeV/$c^{2}$}
\def \pgev     {GeV/$c$}
\def\stilde{\widetilde}
\def \D0  {${\rm D}{\rm O\!\!\!\!\hspace{0.5pt}\raisebox{0.2ex}{/}}$}

\begin{document}

\onecolumn
\title{Search for the Supersymmetric Partner of the Top Quark in Dilepton Events 
from  $\mathbf{p\bar p}$ Collisions at $\mathbf{\sqrt{s} = 1.8}$ TeV
\boldmath }
\maketitle
\font\eightit=cmti8
\def\r#1{\ignorespaces $^{#1}$}
\hfilneg
\begin{sloppypar}
\noindent
D.~Acosta,\r {14} T.~Affolder,\r {25} H.~Akimoto,\r {51}
M.~G.~Albrow,\r {13} D.~Ambrose,\r {37}   
D.~Amidei,\r {28} K.~Anikeev,\r {27} J.~Antos,\r 1 
G.~Apollinari,\r {13} T.~Arisawa,\r {51}
% ...
\end{sloppypar}
\vskip .026in
\begin{center}
(CDF Collaboration)
\end{center}

\vskip .026in
\begin{center}
\r 1  {\eightit Institute of Physics, Academia Sinica, Taipei, Taiwan 11529, 
Republic of China} \\
\r 2  {\eightit Argonne National Laboratory, Argonne, Illinois 60439} \\
\r 3  {\eightit Istituto Nazionale di Fisica Nucleare, University of Bologna,
I-40127 Bologna, Italy} \\
\r 4  {\eightit Brandeis University, Waltham, Massachusetts 02254} \\
\r 5  {\eightit University of California at Davis, Davis, California  95616} \\
\r 6  {\eightit University of California at Los Angeles, Los 
Angeles, California  90024} \\ 
\r 7  {\eightit University of California at Santa Barbara, Santa Barbara, California 
93106} \\ 
\r 8 {\eightit Instituto de Fisica de Cantabria, CSIC-University of Cantabria, 
39005 Santander, Spain} \\
\r 9  {\eightit Carnegie Mellon University, Pittsburgh, Pennsylvania  15213} \\
\r {10} {\eightit Enrico Fermi Institute, University of Chicago, Chicago, 
Illinois 60637} \\
\r {11}  {\eightit Joint Institute for Nuclear Research, RU-141980 Dubna, Russia}
\\
\r {12} {\eightit Duke University, Durham, North Carolina  27708} \\
\r {13} {\eightit Fermi National Accelerator Laboratory, Batavia, Illinois 
60510} \\
\r {14} {\eightit University of Florida, Gainesville, Florida  32611} \\
\r {15} {\eightit Laboratori Nazionali di Frascati, Istituto Nazionale di Fisica
               Nucleare, I-00044 Frascati, Italy} \\
\r {16} {\eightit University of Geneva, CH-1211 Geneva 4, Switzerland} \\
\r {17} {\eightit Glasgow University, Glasgow G12 8QQ, United Kingdom}\\
\r {18} {\eightit Harvard University, Cambridge, Massachusetts 02138} \\
\r {19} {\eightit Hiroshima University, Higashi-Hiroshima 724, Japan} \\
\r {20} {\eightit University of Illinois, Urbana, Illinois 61801} \\
\r {21} {\eightit The Johns Hopkins University, Baltimore, Maryland 21218} \\
\r {22} {\eightit Institut f\"{u}r Experimentelle Kernphysik, 
Universit\"{a}t Karlsruhe, 76128 Karlsruhe, Germany} \\
\r {23} {\eightit High Energy Accelerator Research Organization (KEK), Tsukuba, 
Ibaraki 305, Japan} \\
\r {24} {\eightit Center for High Energy Physics: Kyungpook National
University, Taegu 702-701; Seoul National University, Seoul 151-742; and
SungKyunKwan University, Suwon 440-746; Korea} \\
\r {25} {\eightit Ernest Orlando Lawrence Berkeley National Laboratory, 
Berkeley, California 94720} \\
\r {26} {\eightit University College London, London WC1E 6BT, United Kingdom} \\
\r {27} {\eightit Massachusetts Institute of Technology, Cambridge,
Massachusetts  02139} \\   
\r {28} {\eightit University of Michigan, Ann Arbor, Michigan 48109} \\
\r {29} {\eightit Michigan State University, East Lansing, Michigan  48824} \\
\r {30} {\eightit Institution for Theoretical and Experimental Physics, ITEP,
Moscow 117259, Russia} \\
\r {31} {\eightit University of New Mexico, Albuquerque, New Mexico 87131} \\
\r {32} {\eightit Northwestern University, Evanston, Illinois  60208} \\
\r {33} {\eightit The Ohio State University, Columbus, Ohio  43210} \\
\r {34} {\eightit Osaka City University, Osaka 588, Japan} \\
\r {35} {\eightit University of Oxford, Oxford OX1 3RH, United Kingdom} \\
\r {36} {\eightit Universita di Padova, Istituto Nazionale di Fisica 
          Nucleare, Sezione di Padova, I-35131 Padova, Italy} \\
\r {37} {\eightit University of Pennsylvania, Philadelphia, 
        Pennsylvania 19104} \\   
\r {38} {\eightit Istituto Nazionale di Fisica Nucleare, University and Scuola
               Normale Superiore of Pisa, I-56100 Pisa, Italy} \\
\r {39} {\eightit University of Pittsburgh, Pittsburgh, Pennsylvania 15260} \\
\r {40} {\eightit Purdue University, West Lafayette, Indiana 47907} \\
\r {41} {\eightit University of Rochester, Rochester, New York 14627} \\
\r {42} {\eightit Rockefeller University, New York, New York 10021} \\
\r {43} {\eightit Instituto Nazionale de Fisica Nucleare, Sezione di Roma,
University di Roma I, ``La Sapienza," I-00185 Roma, Italy}\\
\r {44} {\eightit Rutgers University, Piscataway, New Jersey 08855} \\
\r {45} {\eightit Texas A\&M University, College Station, Texas 77843} \\
\r {46} {\eightit Texas Tech University, Lubbock, Texas 79409} \\
\r {47} {\eightit Institute of Particle Physics, University of Toronto, Toronto
M5S 1A7, Canada} \\
\r {48} {\eightit Istituto Nazionale di Fisica Nucleare, University of Trieste/\
Udine, Italy} \\
\r {49} {\eightit University of Tsukuba, Tsukuba, Ibaraki 305, Japan} \\
\r {50} {\eightit Tufts University, Medford, Massachusetts 02155} \\
\r {51} {\eightit Waseda University, Tokyo 169, Japan} \\
\r {52} {\eightit University of Wisconsin, Madison, Wisconsin 53706} \\
\r {53} {\eightit Yale University, New Haven, Connecticut 06520} \\
\end{center}
\vspace{2.0ex}
\begin{center}
(Submitted to Physical Review Letters)
\end{center}
\begin{abstract}
\par We have searched for the supersymmetric partner of the top quark (stop) 
in 107 \invpb of $p\bar p$ collisions 
at 
$\sqrt{s}$= 1.8 TeV collected by the Collider Detector at Fermilab (CDF). 
Within the framework of the 
Minimal Supersymmetric extension of the Standard Model (MSSM) each of the 
pair-produced stops is assumed to decay  into 
a lepton, bottom quark and supersymmetric neutrino. Such a scenario would 
give rise to events with 
two leptons, two hadronic jets, and a substantial imbalance of transverse 
energy. No evidence 
of such a stop signal has been found. We calculate a 95\% 
confidence level (C.L.) upper limit on the stop production cross section,  
which excludes stop masses in 
the region ($80\leq m_{\stop}\leq 135$~\mgev) in the mass plane of stop 
versus sneutrino.\\[2.0ex]
\end{abstract}


\twocolumn
One of the 
most promising theories beyond the Standard Model (SM) \cite{sm1} is 
the Minimal Supersymmetric Standard Model (MSSM) \cite{mssm1}. It 
predicts that 
each SM particle has a superpartner (sparticle) with the same quantum numbers, 
except for spin which differs by one half unit.   
Experimental results indicate that supersymmetric (SUSY) particles are 
generally 
not as light as their 
SM partners. SUSY, therefore,
is broken at or above  the electroweak scale, and we treat the
sparticle masses as free parameters.
Due to the large top quark mass, there may be a 
 large mixing between the superpartners of the left and right helicity states 
of the top quark
\cite{stop_mass}. This can lead to substantial mass splitting 
of the stop mass eigenstates ($\tilde{t}_1$, $\tilde{t}_2$) with the 
lighter one 
 (denoted $\tilde{t}$ from now on) potentially  
being the lightest squark. 

 Stop-antistop  pairs ($\stop\bar\stop$) 
 are strongly produced in the $p\bar p$ collisions at the Fermilab Tevatron 
if kinematically accessible. The production cross section has been calculated  
 using QCD in the next-to-leading order (NLO) approximation \cite{stop_xsection}. 
For a given stop mass ($m_{\stop}$) the cross section depends only weakly 
on the other parameters of the MSSM.
In the mass region of interest to our search ($m_{\stop}$=80-140~\mgev),  
the cross section drops from  44~pb to  1~pb. 

 We assume SUSY $R$-parity \cite{r_parity1} conservation, from which 
the stability of the lightest supersymmetric particle (LSP) follows. 
All SUSY particles, including the stop, eventually decay into this LSP.
 Stop decays into the top quark are kinematically not accessible in our 
region of interest due to the high 
top mass ($m_{\stop}<m_t$). 
For the  stop decay into a bottom quark and an on-shell 
chargino ($\stilde \chi_1^\pm$),  
only a very small 
window of opportunity remains at the Tevatron due to the high 
$\stilde \chi_1^\pm$ mass limit from LEP2 \cite{lep_chargino_limit}.
Another possible 2-body stop decay would be the flavor-changing, 
$\stilde t \rightarrow c\tilde\chi^0_1$, decay \cite{stop_to_charm}. 
It would 
proceed via higher order loop diagrams and is thus highly suppressed.    
The 3-body decay into a charged supersymmetric lepton, 
$\tilde t \rightarrow \ \stilde l \nu b$, is closed  
 for most of the stop region currently within the reach of 
CDF because of the slepton mass 
limit of LEP2 \cite{lep_chargino_limit}. 
The existing mass limit of the supersymmetric neutrino, 
$m_{\tilde\nu}\geq 45$~\mgev ~\cite{sneutrino_limit}, 
leaves the decay 
into sneutrino,  $\tilde t \rightarrow \ l \stilde\nu b$, 
open. This 3-body decay proceeds via a virtual chargino, 
and is expected to yield  
equal $e,\ \mu$, and $\ \tau$ branching ratios.

Stop pair production with the  $\tilde t \rightarrow \ l \stilde\nu b$  
decay will yield two 
leptons with opposite electric charge, 
two hadronic jets from the bottom quarks and  considerable 
transverse energy imbalance ($\MET$)  in the 
detector \cite{eta_and_cdf_coordinates} due to the escaping sneutrinos. 
CDF has reported earlier on an analysis based on $B$ 
identification \cite{niki_b_l_neutrino}. In this Letter
we use dilepton events.
Only a few SM 
processes  yield dileptons and can thus potentially mimic our stop 
signature. The most significant  
ones are (1) $t\bar t$ production with leptons from $W$ and/or 
bottom decays; (2) heavy flavor, {\it i.e.}\  
 $b\bar b$ and $c\bar c$ with semileptonic decays; (3) Drell-Yan production 
with hadronic jets from higher
 order processes; (4) diboson production, $WW$, $W Z$ and  $ZZ$; (5) lepton 
pairs from the decay of vector mesons, 
such as 
$J/\psi$ and $\Upsilon$; (6) events without two genuine prompt leptons, 
where a hadron is misidentified as a lepton, or 
decays in flight to a lepton.

The search presented here is based on 107 \invpb of $p\bar p$ collisions 
at $\sqrt s$ = 1.8 TeV 
collected by the Collider Detector at Fermilab (CDF) during the 
1992 to 1995 running period of the Tevatron.
 A detailed description of the CDF detector can be found in 
Ref.\ \cite{cdfdetector}.
Online triggers selected approximately 6.4 million single lepton 
events and an additional 3.3 million dilepton 
events.
All of those events have been reconstructed, and 13,295 events were 
selected as a  dilepton sample, by requiring  at 
least one tight electron ($\ET\geq 10$~GeV, $|\eta|\leq 1.0$) 
or muon (${\rm p}_{\rm T}\geq 10$~\pgev, $|\eta|\leq 0.6$) candidate, 
and a second  
loose electron ($\ET\geq 6$~GeV, $|\eta|\leq 1.0$) 
or muon (${\rm p}_{\rm T}\geq 6$~\pgev, $|\eta|\leq 1.0$) candidate. 
No explicit tau lepton identification was done, but taus can enter the 
search sample 
if they decay leptonically. 
Electrons are identified by energy deposition in the electromagnetic 
calorimeter with a 
track of corresponding energy in the central drift chamber (CTC) 
pointing to it.
Muons are identified by track segments in both the CTC and the muon 
drift chambers that are located 
behind 4.5 to 10 interaction lengths
of absorber. 
Standard lepton identification cuts are used and 
described elsewhere \cite{top_prd}.
Each lepton is required to  be  isolated, {\it i.e.}\  we require 
the total ${\rm p}_{\rm T}$ of all other tracks within a  
cone $\Delta R\equiv
\sqrt{(\Delta\eta)^2 + (\Delta\phi)^2}\leq 0.4$ around the 
lepton's 
track not to exceed 4~\pgev. 
The jets were reconstructed with a cone algorithm with cone 
radius $\Delta R = 0.7$ \cite{jet_algo}.
We require at least one jet in the central region of the 
calorimeter ($|\eta|\leq 1.0$) 
with $\ET\geq$ 15~GeV, that is separated by 
$\Delta R\geq 0.7$ from both leptons in the event. 
For increased efficiency, we require only one of the two jets to 
be identified. 
Sequential $B$ decays, $J/\psi$, $\Upsilon$ and $Z$ events were 
removed requiring the invariant dilepton masses 
$m_{ll'} \geq 6$~\mgev ~or  $m_{ll} \geq 12$~\mgev ~and  
excluding $76 < m_{ll}<106$~\mgev~(where prime indicates 
any mixture of $e$ and $\mu$ flavors and no prime 
indicates same-flavor dileptons). At the preselection level we 
start 
with $\MET\geq 15$~GeV. Experimental backgrounds, like electrons 
from conversions and muons from 
cosmic rays are removed with additional cuts \cite{arnoldthesis}.
176 events fulfill the above preselection requirements. 

To  estimate the 
number of SM and stop  events in the sample, events of the various 
physics processes
are generated by ISAJET \cite{isajet}  
and simulated for the CDF detector.
We have used CTEQ-3 
parton distribution functions (PDF) \cite{cteq}. 
The stop production cross section was calculated with 
PROSPINO \cite{prospino} and the 
ISAJET cross section was adjusted accordingly.
We have generated events over a large range of stop 
(80-140~\mgev) and sneutrino (45-90~\mgev) masses.

The Drell-Yan and $t\bar t$ production cross sections 
were normalized to CDF measurements \cite{dyan_ttop_cdf}.
The Monte Carlo (MC) $b\bar b$ and $c\bar c$ cross sections  
were verified by  inclusive electron-muon 
samples \cite{james_verify}. 
The $B^0\bar {B^0}$ oscillation effect was added based on the CDF measured 
inclusive $\chi$ mixing fraction \cite{cdf_b0b0}.
The diboson production cross sections of the MC were scaled  to 
those of  NLO calculations \cite{dibosons}.

For low ${\rm p}_{\rm T}$ leptons the contribution due to 
misidentification can be significant and is calculated in two 
steps \cite{arnoldthesis}. First we measure in various data 
samples the so-called ``fake lepton probabilities'' 
(momentum-dependent, separately for electrons and muons, and 
dependent on detector region).  
These ``fake lepton probabilities'' include  
hadrons being misidentified as electrons or muons, and also 
include leptons from in-flight decays 
of pions and kaons.
The data samples used were (i) a minimum-bias trigger sample 
and (ii) a sample with a 50~GeV jet trigger threshold. We 
remove events in those samples 
similarly to the dilepton analysis, and, to be unbiased, we 
exclude all hadrons associated with a jet required 
by the online trigger. 
We measure misidentification probabilities between 0.4\% and 7\% 
 for both $e$'s and $\mu$'s \cite{arnoldthesis}, and the 
results are consistent between the two data samples.

Second, in a single lepton sample  
we use these ``fake lepton probabilities'' successively on 
each other track in the event 
to simulate dilepton events. 
We use the ``fake lepton probabilities'' to simulate both the 
number of misidentified-lepton events as well as their kinematic 
properties.

\begin{figure}[tbh!]
\begin{center}
 \textbf{CDF}\ \ \ \ \ \ \  \boldmath$\int {\rm L dt}=\,$
\unboldmath\textbf{107}\boldmath${\rm pb}^{\mathbf{-1}}$
\unboldmath\\
\epsfig{file=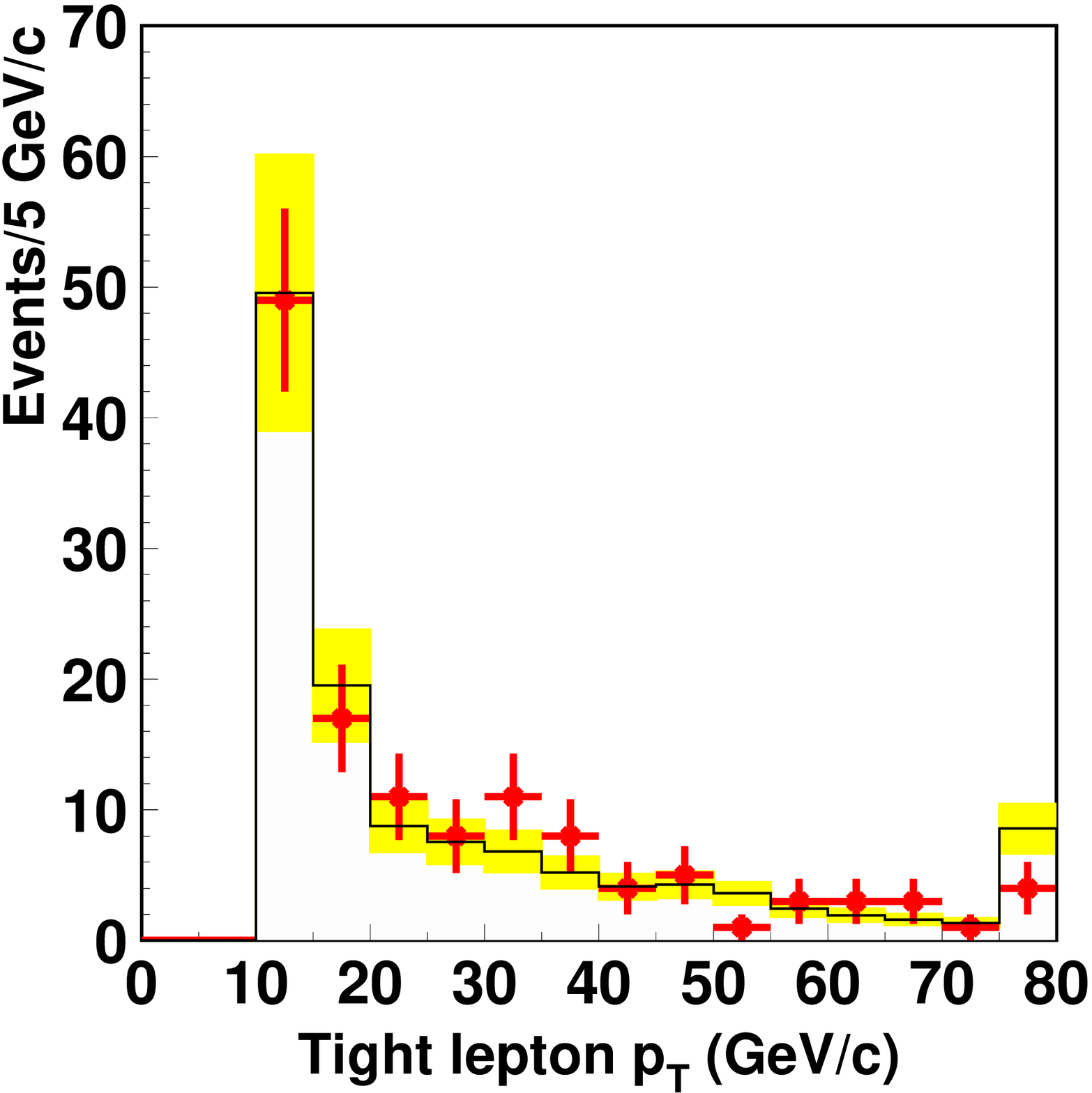,width=1.55in}
\epsfig{file=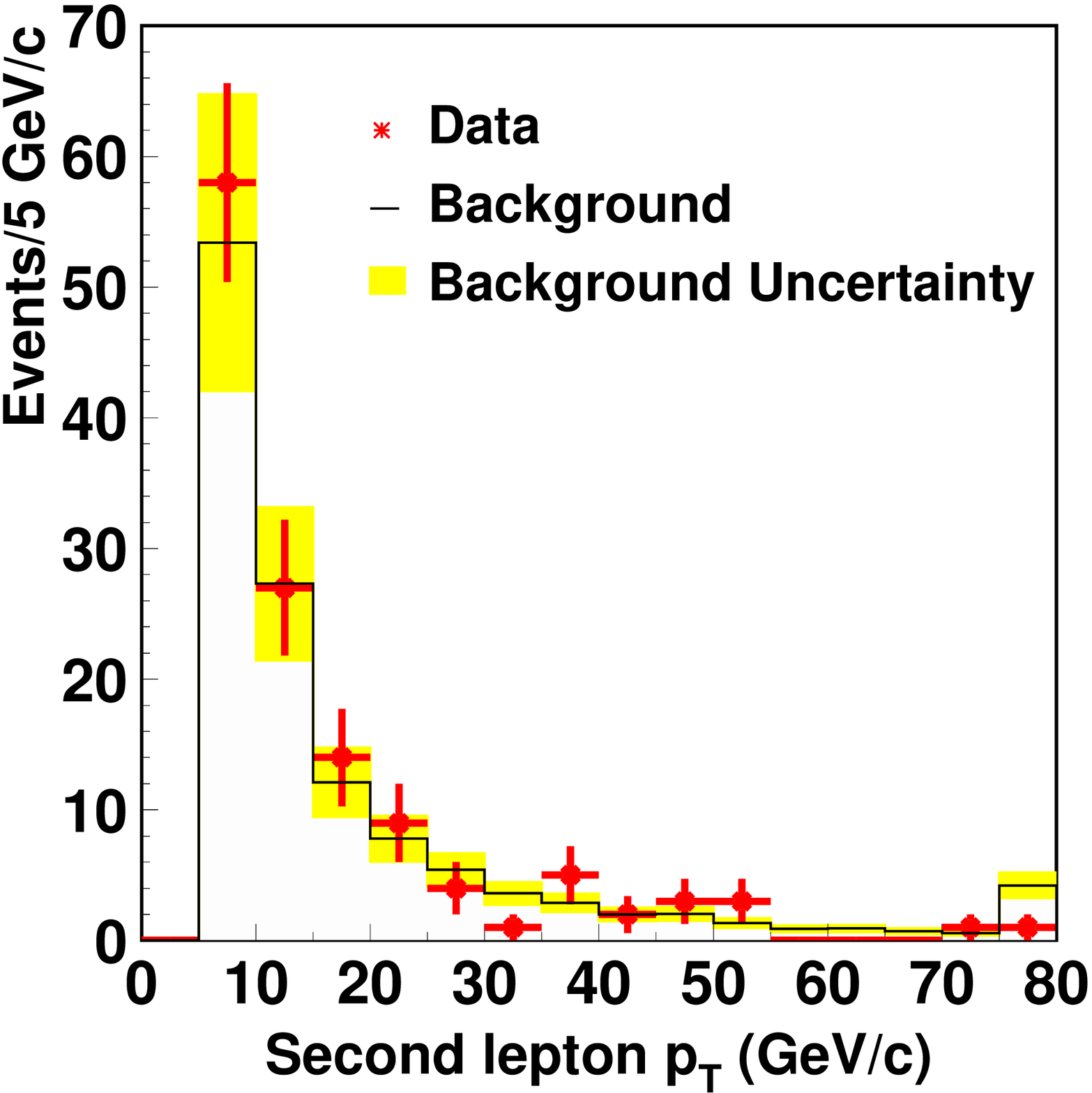,width=1.55in}\\[0.5pc]
\epsfig{file=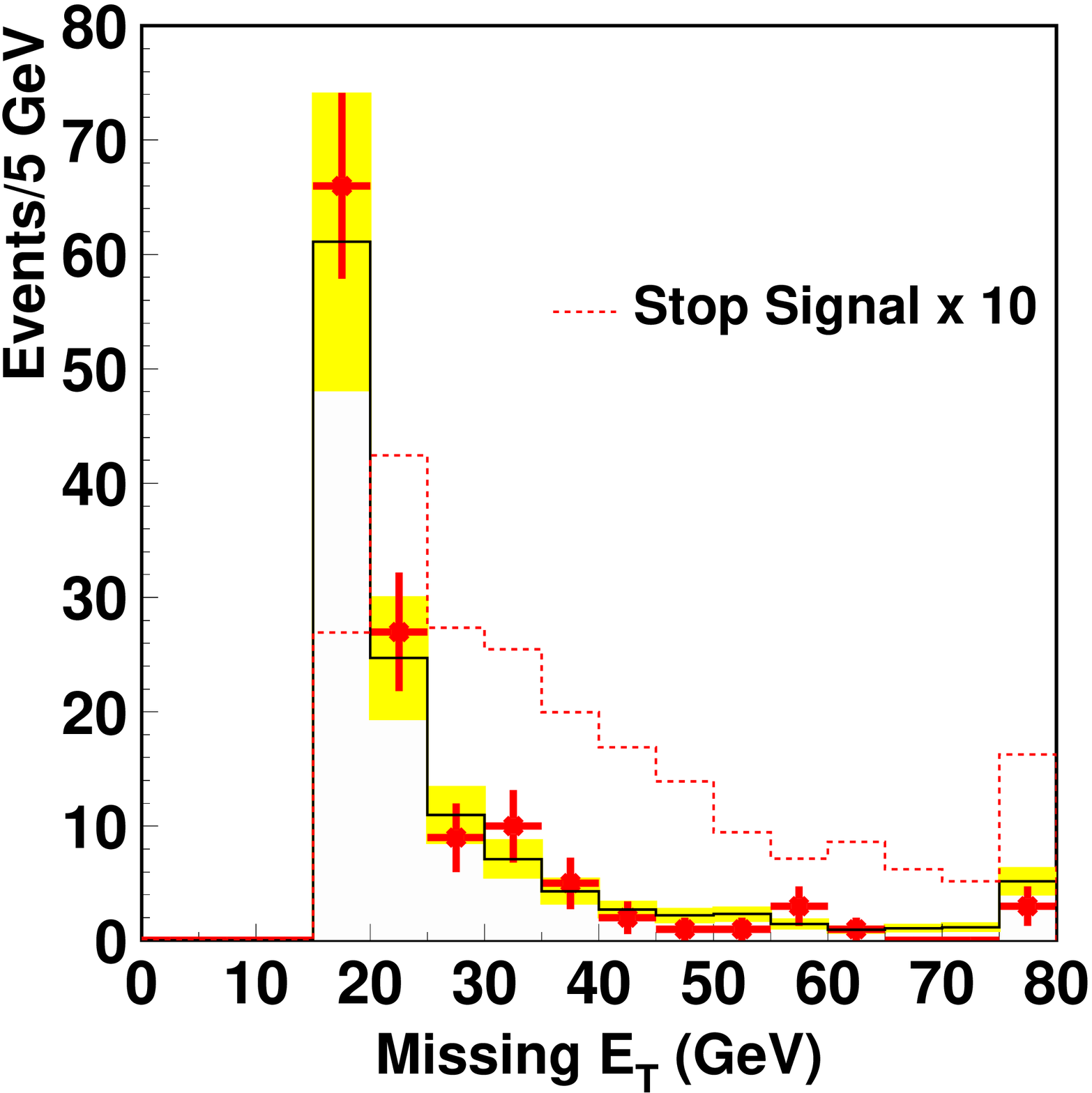,width=1.55in}
\epsfig{file=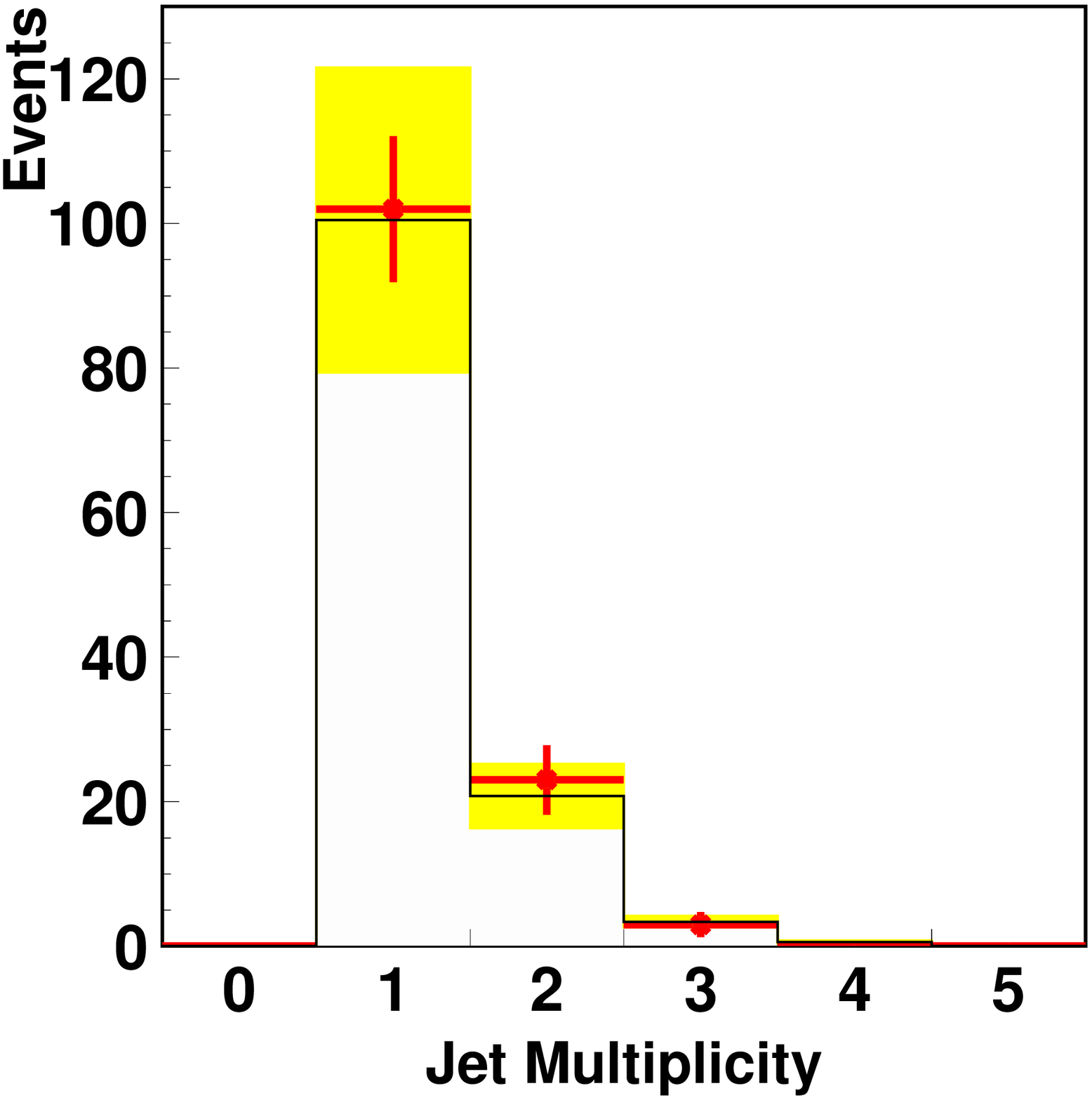,width=1.55in}
\caption{Data and expected  background after preselection. 
Tight and second lepton transverse energies, missing 
transverse energy (for comparison we also show 
 the missing  $E_{\rm T}$ distribution for a 10 times 
stop signal of 
  $m_{\tilde t}$ ($ m_{\tilde\nu}$) = 100 (75)~GeV/$c^2$) and  
jet multiplicity 
distributions shown for  events with opposite charge leptons. 
The last high bins contain overflows. }
\label{fig:presel_data_vs_back}
\end{center}
\end{figure}

The major background to the preselection sample comes from 
heavy flavor production, with about a quarter of the events 
having leptons of the same charge. Another significant background 
comes from 
Drell-Yan processes. In those events the $\MET$ comes from $\tau$ 
decays 
or jet and lepton energy mismeasurements due to uninstrumented 
detector regions. 
We expect a total background of 155 $\pm$ 55
events, while a stop and sneutrino mass combination of 
100 and 75~\mgev ~would contribute 24 $\pm$ 9 events.  
Table~\ref{tab:after_presel} shows the expected contributions 
in detail for like sign and opposite sign leptons. 
To verify our background calculation further, we compare kinematic 
distributions of the data and 
the expected background. Figure~\ref{fig:presel_data_vs_back} shows a 
few such distributions. Top and diboson 
production yield generally more energetic leptons than  
$ b\bar{b}$, $ c\bar{c}$ or misidentified leptons.  
The ${\rm p}_{\rm T}$ distributions of the leptons show 
that both high and low ${\rm p}_{\rm T}$ lepton sources agree well 
with the observed 
data. The observed  $\MET$ distribution agrees both at low $\MET$, 
where detector effects dominate, 
and at high $\MET$, where neutrinos from $W$ and $Z$ bosons determine 
the spectrum.
We also note  
that the parton shower MC  describes well the observed jet multiplicity. 
From the signal to background ratio, it is clear that the preselection 
sample does not have
 sufficient sensitivity to answer the question of stop pair production. 
In contrast  
 to an earlier search \cite{d0} we select a kinematic region in which
we expect higher S/B where S is the stop signal and B is the 
background which includes 
SM processes and misidentified leptons. This decreases the 
systematic uncertainty 
associated with 
low ${\rm p}_{\rm T}$ leptons in our analysis. 
\begin{table}[tbh!]
\caption{Data, expected backgrounds for the pre\-se\-lection sample, and 
expected stop signal for $m_{\tilde t}$ ($ m_{\tilde\nu}$) = 100 (75)~\mgev. 
The stop event acceptance is 2.5\% 
 at this stage. 
\label{tab:after_presel}}
\begin{center}
\begin{tabular}{lr@{.}l@{~}c@{~}r@{.}lr@{.}l@{~}c@{~}r@{.}l}
Source&\multicolumn{5}{c}{OS} &\multicolumn{5}{c}{LS} \\
\hline 
Drell-Yan                      &  52 & 2  & $\pm$ &   13 & 7   &  0 & 4     
& $\pm$ &     0 & 4  \\
\( b\bar{b} \), \( c\bar{c} \) &  43 & 5  & $\pm$ &   32 & 1   & 16 & 4     
& $\pm$ &    17 & 6  \\
\( t\bar{t} \)                 &   9 & 5  & $\pm$ &    2 & 9   &  0 & 6     
& $\pm$ &     0 & 2  \\
$WW,WZ,ZZ$                     &   3 & 8  & $\pm$ &    0 & 9   &  0 & 4     
& $\pm$ &     0 & 1  \\
Misidentified Leptons          &  16 & 3  & $\pm$ &    4 & 4   & 12 & 4     
& $\pm$ &     3 & 4  \\
\hline 
Total Background               & 125 & 2  & $\pm$ &   46 & 7   & 30 & 1     
& $\pm$ &    18 & 4  \\
Data &  \multicolumn{5}{c}{128}&  \multicolumn{5}{c}{48}     \\
Expected  $\stop\bar\stop$     &  22 & 6  & $\pm$ &    8 & 9   & 1 & 0      
& $\pm$ &     0 & 4  \\
\end{tabular}
\end{center}
\end{table}

In less than 5\%
of stop events the two leptons will have the same electric charge due to the 
semileptonic decay of one of the $b$-quarks.
However, 20\% 
of the SM background yields like-sign (LS) lepton events. 
We thus focus our search on events with opposite-sign charge (OS) 
leptons.
For $R_p$-conserving supersymmetry we expect large missing energy 
from the rather heavy sneutrinos. 
In Fig.~\ref{fig:presel_data_vs_back} we see most of the background 
events clustering at low missing $\ET$. 
A  $\MET$ cut of 30~GeV removes 77\%
of the SM background but keeps about 57\%
of the stop events. Energy mismeasurement of leptons, or the presence 
of neutrinos from Drell-Yan $\tau$ decays, would cause the leptons 
(and the dilepton system as well) 
to be aligned with the $\MET$ direction. This is not typical for the signal, 
where we expect true $\MET$ 
from the sneutrinos in the stop decay. 
We therefore require the azimuthal angles between $\MET$ and the individual 
leptons and the dilepton 
system ${\Delta \phi}^{\sMET}_{l_1}$, ${\Delta \phi}^{\sMET}_{l_2}$, and 
${\Delta \phi}^{\sMET}_{l_1l_2}$ to be larger than $30^{\circ}$.

In Drell-Yan plus jets events or when $b\bar b$ or $c\bar c$ 
events originate from 
gluon splitting (initial or final state) events, the two   
leptons balance  the jets in the transverse plane. We veto events    
 where the angle between either lepton and the most energetic 
central jet, 
${\Delta \phi}^{\rm jet}_{l_1}\ {\rm and}\ {\Delta \phi}^{\rm jet}_{l_2}$,
  is larger than $90^{\circ}$. 

Events from top pair production pass the above cuts 
with efficiencies similar to stop pair events and are now the dominant 
source of SM background. In top events the leptons 
come from $W$ decay and are very energetic. In the case of stop, we 
have 3-body decays containing 
a very heavy sneutrino. The amount of available energy in the decay 
depends on the stop-sneutrino mass difference, 
 $\Delta m_{\stop-\tilde\nu}$. For small mass difference, the leptons 
and jets are quite soft and 
 a large fraction of the event energy escapes detection through the 
sneutrinos, unlike a $t\bar t$ event.
For best stop sensitivity at small $\Delta m_{\stop-\tilde\nu}$ we 
require the scalar sum of lepton ${\rm p}_{\rm T}$, 
${\rm p}_{\rm T}^{l_1}+{\rm p}_{\rm T}^{l_2}\leq 75$~\pgev, and the 
${\rm p}_{\rm T}$ of the dilepton system, 
 ${\rm p}_{\rm T}^{l_1l_2}\leq 30$~\pgev. Although a large amount of 
energy escapes undetected, 
the sneutrinos tend to be back-to-back, thus reducing the measured 
$\MET$. We also require the 
 sum of the most energetic central jet $\ET$ and the missing $\ET$, 
$\ET^{\rm jet}+\MET\leq 160$~GeV. 

For large stop-sneutrino mass difference, the 
leptons are more energetic and we can increase our lepton ${\rm p}_{\rm T}$ 
requirement 
to 10~\pgev ~without much loss in stop efficiency. However, 
 leptons and jets are still significantly softer than in $t \bar t$ events. 
We place the same jet, missing $\ET$, and lepton requirements as at 
small $\Delta m_{\stop-\tilde\nu}$, $\ET^{\rm jet}+\MET\leq 160$~GeV and 
${\rm p}_{\rm T}^{l_1}+{\rm p}_{\rm T}^{l_2}\leq 75$~\pgev, but 
loosen the requirement on the dilepton ${\rm p}_{\rm T}$ to 
${\rm p}_{\rm T}^{l_1l_2}\leq 55$~\pgev.

Table~\ref{tab:reduction} shows the expected number of stop events for 
the two search regions.
We start our search at stop masses of 80~\mgev ~to overlap with previous 
LEP limits. Near the kinematic limit
 of the stop decay, $m_{\stop}= m_{\tilde\nu}+m_b$, lepton and jet 
energies become very soft, 
limiting our stop detection capabilities. At high stop mass our 
sensitivity is limited by the steeply 
falling production cross section. 
In the region of interest to this search the final stop event 
acceptance varies 
between 0.3\%
 and 2.3\%.

The biggest source of uncertainty on the number of expected stop 
events arises from the choice of the renormalization and factorization scale, 
$Q^2$, which characterizes the amount of energy transferred  
during the collision.  
The $\MET$ is reduced (due to the sneutrinos being more back-to-back) 
when $Q$ is increased, and the jet $\ET$ gets softer
 when $Q$ is decreased. By varying $Q$ by a factor of two up and down, 
we determine the uncertainty due to the 
 choice of $Q^2$ to be  
 32\%. 
Other significant 
sources of uncertainty are: 
    (1) the choice of PDF (11\%)
    (2) the absolute energy scale of the detector (11\%)
    (3) the amount of gluon radiation (7\%)
    (4) trigger, lepton and isolation efficiency  (5\%),  
and (5) the luminosity measurement (4\%).
The statistical uncertainties of the MC samples are about 8\%.
 Combining the statistical and systematic uncertainties we obtained a 
total uncertainty of 38\% 
for the signal expectation. Similarly, we evaluated  the uncertainty of 
the background calculation 
to be 
 30\%. 
\begin{table}[tbh!]
\caption{Data, expected background, and expected stop signals after final cuts. 
Stop A scenario represents a small $\Delta m_{\stop-\tilde\nu}$ with 
$m_{\tilde t}$ ($ m_{\tilde\nu}$) = 100 (75)~GeV/$c^2$. 
Stop B scenario represents a large  $\Delta m_{\stop-\tilde\nu}$ with 
$m_{\tilde t}$ ($ m_{\tilde\nu}$) = 120 (60)~GeV/$c^2$. 
 \label{tab:reduction}}
\begin{center}
\begin{tabular}{lcr@{.}l@{~}c@{~}r@{.}lr@{}l@{~}c@{~}r@{}lr@{}l@{~}c@{~}r@{}l}
Selection& Data& \multicolumn{5}{c}{Background} &\multicolumn{5}{c}{Stop A} 
&\multicolumn{5}{c}{Stop B} \\
\hline 
Preselection & 176 &  155 & 3  & $\pm$ &  50 & 2   &  23 & .6     & $\pm$ 
&     8 & .9 &  34 & .5     & $\pm$ &     13 & .0   \\
OS \& $\MET$& 26  & 
           28 & 7       & $\pm$ &       8 & 6   &
           12 & .9     & $\pm$ &    4 & .9  &
           25 & .1     & $\pm$ &    9 & .5   \\
${\Delta \phi}^{{\sMET}}_{l,ll}$ \& ${\Delta \phi}^{\rm jet}_{l}$& 4 & 
            8 & 1       & $\pm$ &      2 & 4   &
            6 & .7     & $\pm$ &       2 & .5 &
           14 & .8     & $\pm$ &       5 & .6  \\
\hline
small ${\Delta m}_{\stop-\tilde \nu}$ & 0 & 
            1 & 5      & $\pm$ &      0 & 5  &
            5 & .7     & $\pm$ &       2 & .1 &
              &       & -     &         &   \\
\hline
large ${\Delta m}_{\stop-\tilde \nu}$ & 0 & 
            2 & 1      & $\pm$ &      0 & 5  &
              &       & -     &         &    &
             8& .2     & $\pm$ &    3 & .1   \\
\hline
                 \multicolumn{7}{l}{ } &\multicolumn{5}{c}{} &\multicolumn{5}{c}{}  \\
\hline
                 \multicolumn{7}{l}{95\% 
C.L. cross section limit:         } &\multicolumn{5}{c}{9.0 pb} 
&\multicolumn{5}{c}{2.2 pb} 
\end{tabular}
\end{center}
\end{table}
\begin{figure}[tbh!]
\begin{center}
\epsfig{file=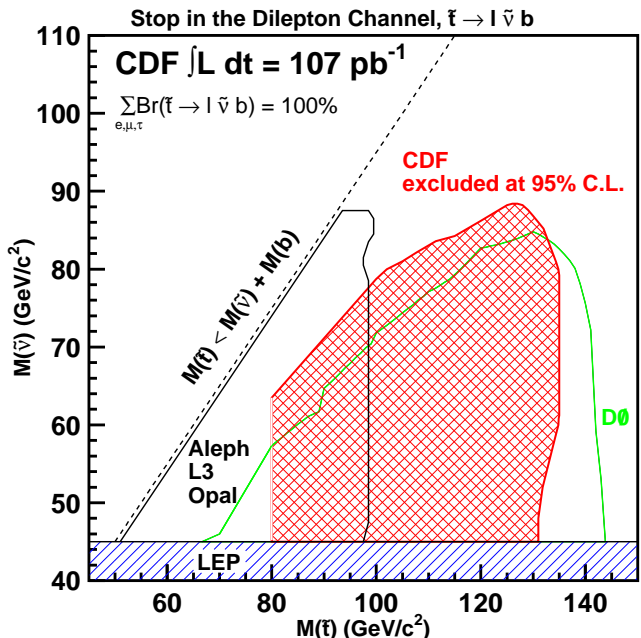,width=3.4in}
\caption{
Stop and sneutrino mass plane showing the CDF 95\% C.L.
excluded region as hatched area.
For the 3-body stop decay, $\tilde t \rightarrow \ l \tilde\nu b$, a 
 33.3\% 
branching ratio to  each of the three lepton flavors is used.}
\label{fig:limit_contour}
\end{center}
\end{figure}

After establishing the selection  cuts by using a ``blind'' analysis technique, 
we then apply the cuts to the preselection data. 
We observe zero events for both the small and the large $\Delta m_{\stop-\tilde\nu}$ 
sets of cuts,  consistent with 
 our background expectation of  1.52$\pm$0.47  and 2.07$\pm$0.46 events. 
 We used 
 the frequentist method \cite{poisson} with zero observed events, no background 
subtraction and  
a total uncertainty of 38\% 
 on the predicted signal to calculate a 95\%
 confidence level upper limit of 4.01 stop events. Consequently, we exclude all 
stop-sneutrino 
 mass combinations that would yield more than 4.01 events. 
Figure~\ref{fig:limit_contour} shows our  
result compared to LEP2 \cite{lep_chargino_limit} and \D0 \cite{d0}. 
The improvement 
 in the small $\Delta m_{\stop-\tilde\nu}$ region 
is due to our 
 increased signal to background ratio of 4:1.

In conclusion, we have searched for stop pair production  in 107 \invpb 
of data 
from $p\bar p$ collisions at $\sqrt{s}=$ 1.8 TeV  collected by CDF. 
The observed dilepton, jet, and missing $\ET$ events are consistent with 
expectations 
from SM sources. Failing to find a signal of supersymmetry, we establish 
mass limits at 95\% C.L.: 
 we exclude stop masses up 
to $m_\stop=135$~\mgev ~(at $m_{\tilde\nu}$ of 72-79~\mgev) and sneutrino 
masses up to 88.4~\mgev
 ~(at $m_\stop$ of 126~\mgev ).  

We thank 
the Fermilab staff and the technical staffs of the participating institutions for their vital contributions.  
This work was supported by the U.S. Department of Energy and National Science Foundation; the Italian Istituto Nazionale di Fisica Nucleare; the Ministry of Education, Culture, Sports, Science, and Technology of Japan; the Natural Sciences and Engineering Research Council of Canada; the National Science Council of the Republic of China; the Swiss National Science Foundation; the A.P. Sloan Foundation; the Bundesministerium f\"{u}r Bildung und Forschung, Germany; and the Korea Science and Engineering Foundation (KoSEF); the Korea Research Foundation; and the Comision Interministerial de Ciencia y Tecnologia, Spain.

\end{document}